\documentstyle[11pt,newpasp,twoside,epsf]{article}
\begin{document}
\title{Newly Discovered Globular Clusters in the Outer Halo of M31} 
\author{A. Huxor, \& N. R. Tanvir}
\affil{University of Hertfordshire, College Lane, Hatfield, Herts, AL10 9AB, UK }
\author{M. Irwin}
\affil{Institute of Astronomy, University of Cambridge, Madingley Road, Cambridge, CB3 0HA, UK}
\author{A. Ferguson}
\affil{Max Planck Institute for Astrophysics, Karl-Schwarzschild-Str. 1, Postfach 1317, D-85741 Garching, Germany}
\author{R. Ibata}
\affil{Observatiore de Strasbourg, 11 rue de l'Universit\'e, F-67000 Strasbourg, France}
\author{G. Lewis}
\affil{School of Physics A28, University of Sydney, NSW 2006, Australia}
\author{T. Bridges}
\affil{Anglo-Australian Observatory, PO Box 296, Epping NSW 1710, Australia}

\begin{abstract}
We present nine newly discovered globular clusters in 
the outer halo of M31, found by a semi-automated 
procedure from a INT Wide Field Camera survey of the region. 
The sample includes a candidate at the yet largest known projected
galactocentric radius from M31.
\end{abstract}

\section{Introduction}
 $\Lambda$CDM cosmological models, and their semi-analytical
extensions, have had considerable success in describing hierarchical
structure formation and galaxy evolution.  However, important
questions still remain. For example, where are the predicted large
number of low mass dark haloes?
What are the profiles of the dark haloes?  How important are major and
minor mergers in building galaxies and star formation?  One
way to address such issues is by detailed study of nearby resolved galaxies,
 to deduce evolutionary history from their present day
 structure.
The galaxies of the local group are ideal for such investigations, and
a wide area CCD survey of the halo of M31 has been
 undertaken by our collaboration.
This survey has already revealed  
previously unknown detail and structure (Ferguson et al. 2002), including 
the discovery of a large stellar stream, 
presumably the relic of dwarf galaxy destruction (Ibata et al. 2001). 
We have now used this survey to complete
 a search for previously
undiscovered globular clusters, which are important tracers of the
dynamics and chemistry of the outer halo . 

\section {GCs as dynamical and chemical tracers}
Recent modelling by Evans et al. (2000) suggests 
that, contrary to accepted wisdom, the
mass of the halo of M31 is comparable to, or less than,
that of the Milky Way. 
However the uncertainties are large.  For
example, test particles at large galactocentric radius are 
important to constrain the total mass, but are currently limited to 
dwarf satellites and other local group galaxies, whose 
past interactions with, for example, Milky Way are not known.  Thus Evans
and Wilkinson (2000) emphasise the need for more  test particles at an
appropriate distance.
Globular clusters (GCs) are particularly useful, 
being relatively bright they enable good 
spectroscopy, and are found far from the host galaxy (e.g. see figure 1), hence our 
motivation in seeking new examples in the outer halo.
Furthermore, there is evidence that 
they occur in distinct sub-populations, as characterised 
by position, velocity, and 
colour (Perrett et al. 2003). These populations 
provide a fossil record of the
formation history of the galaxy.
Various routes have been proposed for the production of
GCs. Each produces
specific signatures being found in that GC population, which
will help to unravel the history of the M31 system. Hence, we are also interested
in the metallicity of our candidate globular clusters, in addition to their
kinematics.

\section{Candidate identification}
Initial candidates 
were found from the survey database, produced by the standard INT
Wide Field Survey pipeline and provided by the Cambridge Astronomical
Survey Unit. Candidate selection, was based on magnitude, colour, ellipticity, 
object classification (stellar/non-stellar) and image width (sigma).
These values were iteratively refined, by ensuring that known good quality cluster 
candidates were accepted by the cuts. 
Few GCs are expected at such large galactocentric radii, and
selecting candidates based solely on the criteria listed above
 would produce too many false positives for spectroscopic
followup. 
Hence the initial list of candidates were visually inspected to identify final candidates,
 and these were finally cross-checked against existing catalogues 
(especially the online catalogues maintained by
 Barmby and by Huchra) giving the final nine, previously unknown, GCs presented here. 

\section{Future work}
One of the candidate GCs, at the extreme south-west of the M31 survey field, lies 
at the largest known projected galactocentric radius from M31.
Spectroscopy of this object was obtained with WHT/ISIS
in service time at the end of 2002. The data is currently being reduced, 
and the results will presented in a later paper. Further time for radial velocity 
and metallicity determination
of the other eight candidates is currently being sought.

\begin{figure}[ht]
\plotfiddle{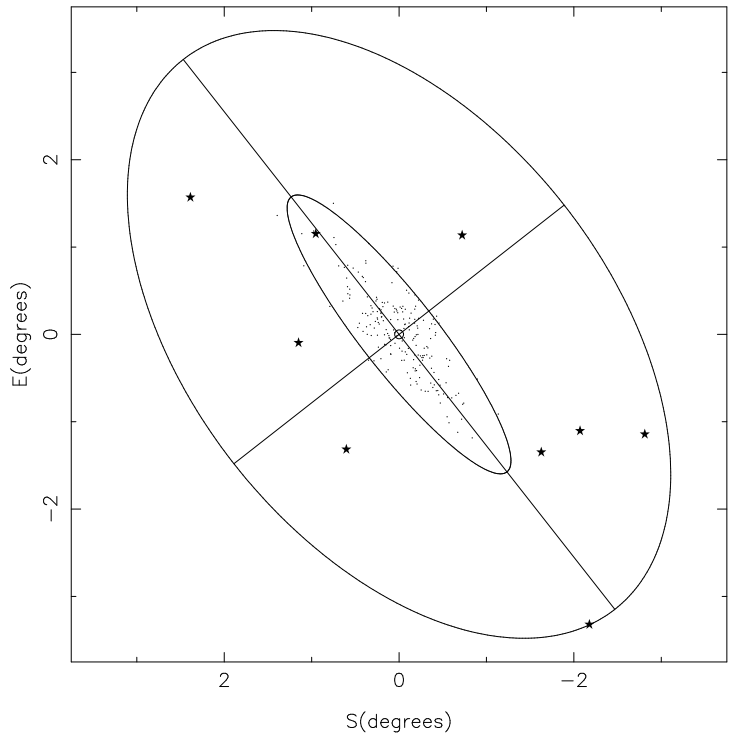}{11cm}{270}{55}{55}{-225}{335}
\caption{The area of our imaging survey (roughly outlined by the large ellipse)
in relation to the visual extent of the M31 disk (small ellipse).
The 
new GCs (asterisks in figure) contribute additional dynamical probes which, 
although few in 
number, extend well beyond the region for which radial velocities have 
been published (points in figure) by Perrett et al. (2002).
}
\end{figure}

\begin{figure}
\plotfiddle{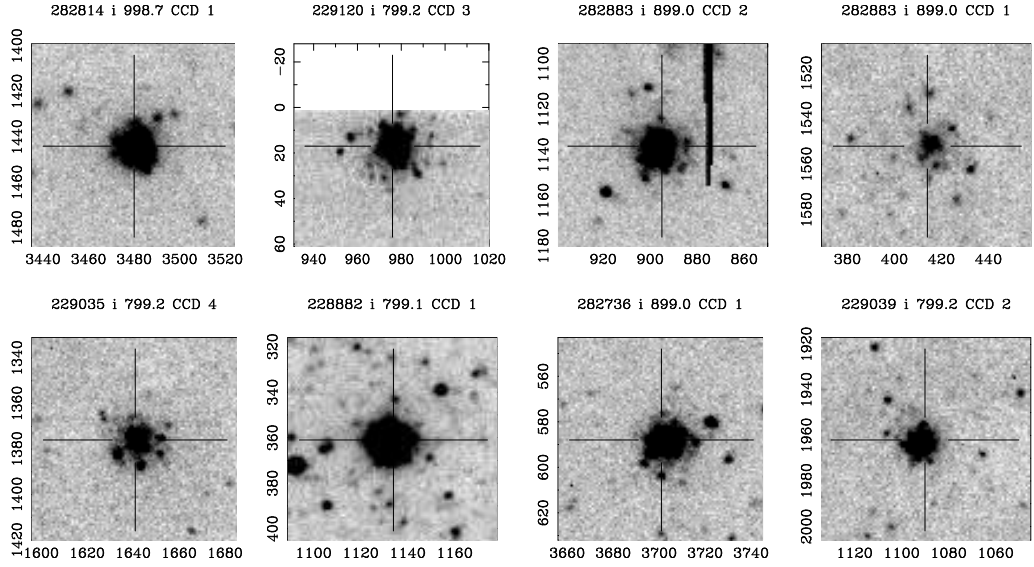}{8cm}{270}{50}{50}{-215}{280}
\caption{Eight of the globular cluster candidates. Images are from
the INT Wide Field Camera survey, taken in i-band, and each is 30" x 30". North
is up and East is left.}
\end{figure}

\begin{figure}
\plotfiddle{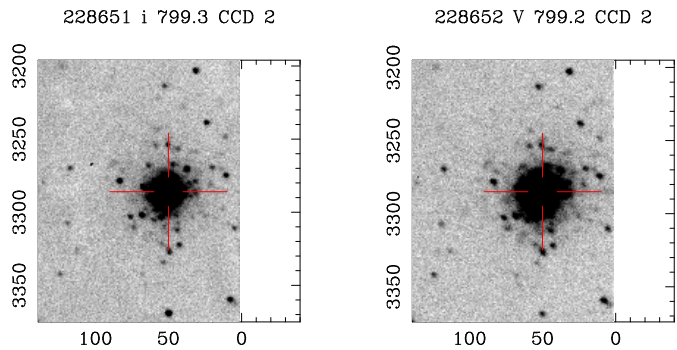}{5cm}{270}{50}{50}{-150}{155}
\caption{i and V band images of the GC candidate that is at the 
largest known project galactocentric distance, some 55 kpc. Images also from the INT/WFC
survey, and have same size and orientation as in figure 2.}
\end{figure}


\begin{references} 
\reference Evans, N. W. et al. 2000, \apj, 540, L9
\reference Evans, N. W. \& Wilkinson, M. I. 2000, \mnras, 316, 929
\reference Ferguson, A. M. N. et al. 2002, \aj, 124, 1452
\reference Ibata, R. et al. 2001, Nature, 412, 49
\reference Perrett, K. M. et al. 2003, \apj, 589, 790
\reference Perrett, K. M. et al. 2002, \aj, 123, 2490
\end{references}
\end{document}